\documentclass[prb,amsmath,twocolumn,showkeys,showpacs]{revtex4}
\usepackage{graphicx}
\usepackage{dcolumn}
\usepackage{amssymb}
\usepackage{bm}
\begin{document}
\newcommand{\be}{\begin{equation}}
\newcommand{\ee}{\end{equation}}
\newcommand{\vk}{\mathbf{k}}
\newcommand{\vq}{\mathbf{q}}
\newcommand{\vp}{\mathbf{p}}
\newcommand{\vecr}{\mathbf{r}}
\newcommand{\vx}{\mathbf{x}}
\newcommand{\caH}{\mathcal{H}}
\newcommand{\la}{\langle}
\newcommand{\ra}{\rangle}
\newcommand{\e}{\epsilon}
\title{ Collective excitation of  quantum wires
and  effect of
spin-orbit coupling in the presence of  a  magnetic field along the wire}
\author{Hyun C. Lee}
\affiliation{Department of Physics and Basic Science Research Institute,
 Sogang University, Seoul, Korea}
\author{S. -R. Eric  Yang\footnote{corresponding author,  eyang@venus.korea.ac.kr}}
\affiliation{Department of Physics, Korea University, Seoul, Korea}
\affiliation{School of Physics, Korea Institute for Advanced Study, Seoul, Korea}
\date{\today}
\begin{abstract}
The band structure of  a quantum wire with the Rashba spin-orbit coupling 
develops a pseudogap in the presence of a  magnetic field along
the wire. In such a system spin mixing at the Fermi wavevectors $-k_F$ and $k_F$ can be different. 
We have investigated theoretically the collective mode of this system, and found 
that the velocity of this collective excitation depends  sensitively on the strength of the Rashba spin-orbit interaction 
and magnetic field.  
Our result suggests that the strength of the spin-orbit interaction can be determined
from the measurement of 
the velocity. 
\end{abstract}
\maketitle
\section{Introduction}

\begin{figure}[ht]
\begin{center}
\includegraphics[angle=0,width=0.8\linewidth]{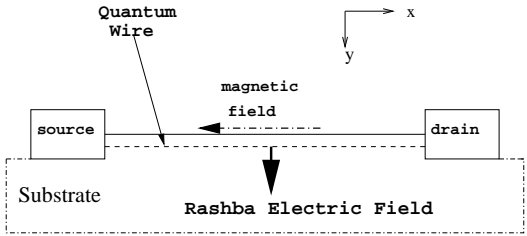}
\caption{The geometry of  a quantum wire with a magnetic field along the wire.}
\label{modulator}
\end{center}
\end{figure}

\begin{figure}[ht]
\begin{center}
\includegraphics[angle=0,width=0.9\linewidth]{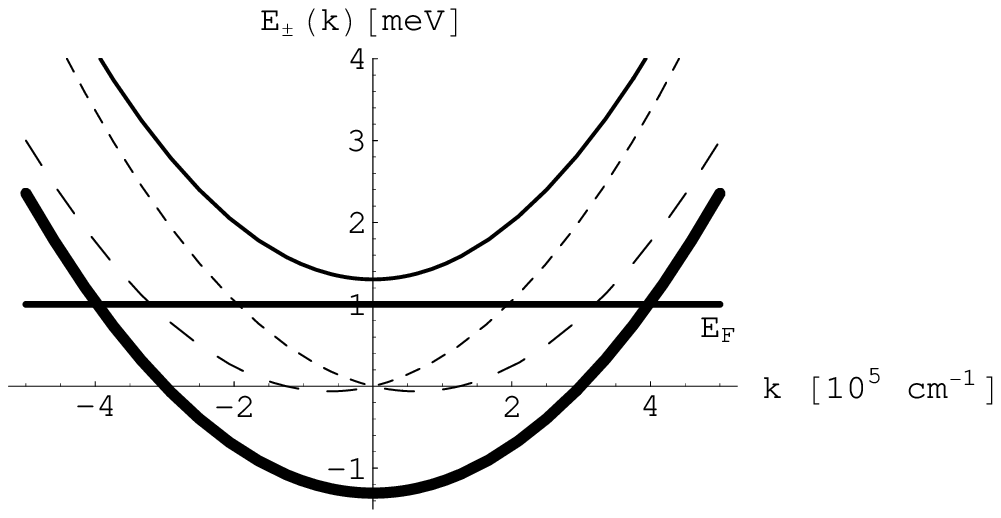}
\includegraphics[angle=0,width=0.9\linewidth]{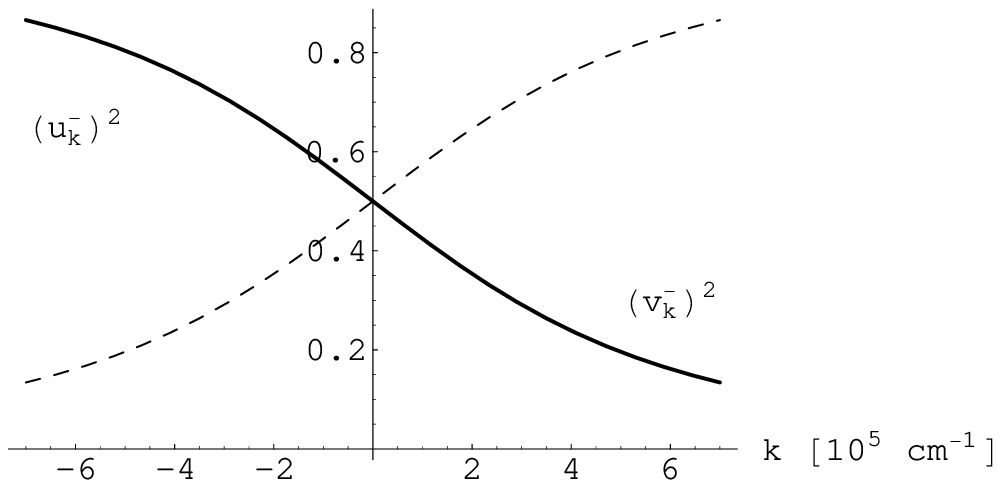}
\caption{
Upper figure: Solid lines represent the lowest energy subband structure of the quantum wire in the absence of the Dresselhaus term.
(Dashed lines are for zero magnetic field).
Note that  the Fermi energy lies in the pseudogap.
When a finite value of magnetic field is present  bands do anticross (In the figure  $B=3 \text{T}$).
The input parameters are  $\eta_R = 2 \times 10^{-9} \text{eV}\cdot \text{cm}$, $m^*=0.024 m_e$.
In this case the numerical value of $g$ is approximately 0.7.
Lower figure: The spin-up (solid line) and -down (dashed line) components  
$(u_k^-)^2$ and $(v_k^-)^2$ for the lower $E_-(k)$ band.
The input parameters are identical with the above figure.
Note that $(v_k^-)^2=1-(u_k^-)^2$.
}
\label{energy}
\end{center}
\end{figure}
Recently active research is taking place on how to manipulate  spin properties of single electrons,
and  several semiconductor spin devices  based on spin-orbit coupling have been proposed.\cite{Rashba2005,wolf2001}
Among these we focus  on 
a spin filter\cite{datta1990}  proposed by St\v reda and \v Seba.\cite{streda2003}
They  proposed a spin filter combining strong Rashba spin-orbit interaction (SOI) and the magnetic field 
\textit{parallel} to a quantum wire (see Fig. \ref{modulator}).
This system has an interesting one-dimensional band structure, ( see Fig.~\ref{energy} ): 
a pseudogap is present at zero wavenumber and   
the orientation of electron spin 
depends on the  wavevector.\cite{streda2003}
For the lower band  the electron  with sufficiently negative $k$ 
is mostly polarized in the +z direction while that of  sufficiently positive $k$ is mostly polarized in the -z direction.
When the Fermi energy lies in the pseudogap  substantial spin-mixing exists  for
moderate value of the Fermi energy.
The transmission/reflection
coefficients of such a wire in the presence of a step potential  has been calculated 
in the presence of electron-electron  interaction using poor man's renormalization group approach.\cite{Dev}

The dispersion of the collective mode of quantum wires in the presence
of a magnetic field perpendicular to the wire has been investigated for many years.\cite{collecB}
Recently the interplay of Rashba SOI and electron-electron interaction 
in quantum wires have been studied by several groups.\cite{yu2004,gritsev2005} 
However, none of these studies have dealt with the case where the applied magnetic field is parallel to quantum wire
in the presence of spin orbit interaction.
In this paper we investigate how the collective electronic properties may be manipulated by spin-orbit coupling.
In II-VI semiconductors
the Rashba term is expected to be larger than the
the Dresselhaus coupling. In III-V semiconductors, such as GaAs, the
opposite is true.\cite{Rashba2005}
However, in these quantum wires  the Dresselhaus term can be rather small
under certain conditions as we argue below.
The   band structure of such quantum wires in the presence of
a parallel magnetic field is as displayed in Fig.~\ref{energy}.
The nature of the collective mode  is unclear when the spin mixing
at the Fermi wavevectors   $-k_F$ and  $k_F$ are different.
We have obtained, employing bosonization methods\cite{voit,bosobook},  
the exact dispersion relation of the collective mode of the lower band
 when the Fermi energy lies in the pseudogap.
The dispersion relation of this mode is  
\be
\label{dispersion}
\omega = \big [v_\theta(q) v_\phi(q)\big ]^{1/2} q \equiv v_o q.
\ee
$v_\theta(q)$ and $v_\phi(q)$ are defined as follows:
\begin{align}
\label{velocities}
v_\theta (q ) &=  v_F \left(1 + \frac{V_q}{\pi v_F}-\frac{g V_{2 k_F}}{2\pi v_F} \right),\nonumber \\
v_\phi (q ) &= v_F \left(1 + \frac{g V_{2 k_F}}{2 \pi v_F} \right ),
\end{align}
where $v_F$ is the Fermi velocity and the renormalization factor of the strength of backscattering is 
\be
\label{g1-coupling}
\begin{split}
g =
 \frac{\e_Z^2}{\e_Z^2+(\eta_R k_F)^2}.
\end{split}
\ee
$\theta$ and $\phi$ are the phase fields which are basically linear combinations of density operators $\rho_{R/L}$ and 
they are  defined in Eq.~(\ref{phase}).
$V_q$ is the interaction matrix element, 
$\e_Z$ is the magnitude of Zeeman coupling, and  $\eta_R$ is a 
parameter characterizing Rashba SOI (see below).
From the expression of $g$ [Eq.~(\ref{g1-coupling}) ]we see that  the velocity of 
this collective excitation depends sensitively on the Rashba SOI and magnetic field.
This result differs from the that of an ordinary Luttinger liquid in that the back scattering term $V_{2 k_F} $ is renormalized
by a factor $g$.  The physical origin of this factor reflects the different spin mixing of single particle 
states near the Fermi wavevectors, which   
are coupled by backscattering. 

The presence of the renormalization factor $g$ may be exploited to determine the constant $\eta_R$.
There is no simple way to calculate $\eta_R$ 
because it depends both on  the electric field inside the semiconductor heterostructure and on
the detailed boundary conditions at the interface. Instead these spin-orbit coupling constants  were measured by
electric, optical, and photoelectrical means.\cite{meas1,meas2,meas3,meas4} 
We suggest that the measurement of the velocity of the collective excitation $v_0$ may provide 
another way to determine the value of $\eta_R$.
This measurement can be carried out using tunneling between two parallel wires in the 
presence of an additional magnetic field $\mathbf{B}_t=\nabla 
\times \mathbf{A}_t$ along the y-axis
\cite{governale2002,boese2001}.
This method allows one to determine the spectrum of elementary excitations\cite{Carpentier,Zulicke}
momenta much larger than $2k_F$ \cite{auslaender2002}.

This paper is organized as follows. In Sec. II we introduce our model and review the results obtained by St\v reda and \v Seba
for the \textit{non-interacting} case. In Sec. III we incorporate the electron-electron interaction and obtain an effective
Hamiltonian for the system. In Sec. IV the dispersion of collective excitation is computed based on the 
effective action obtained in Sec. III. Sec. V we discuss how our result for the velocity differs from 
the results of ordinary Luttinger liquids. An experiment is proposed to measure $\eta_R$.

\section{Model for single particle Hamiltonian }

In our model confinement potentials are present along the y- and z-axis  and 
  quasi-one-dimensional motion of electrons  is possible along the x-axis.
  The widths of the wavefunction along both the 
y- and z-axis are assumed to be negligible.  The lowest subband  energies along the y- and z- axis
are denoted by $E_y$  and $E_z$.
A magnetic field parallel to the quantum wire along x-axis is present $\mathbf{B} = -B  \hat{\mathbf{x}}$. 
The corresponding vector potential can be chosen to be 
 $\mathbf{A} = - B y \hat{\mathbf{z}},\quad B > 0$. 
In our model Rashba electric field is applied along the y-axis
(see Fig. \ref{modulator}), and is given by $ \mathbf{E}=+E_0 \hat{\mathbf{y}},~~(E_0 > 0)$.
The  Rashba
spin-orbit interaction \cite{rashba1,rashba2} then takes the form
\be
\mathcal{H}_R =  
 \eta_R \,\Big(  k_x \sigma_z - k_z \sigma_x  \Big ),
\ee
where  $\eta_R=  |e| \hbar^2 E_0 / 4 m_e^2 c^2 > 0$. 
The strength of Rashba SOI can be controlled by changing electric field.\cite{nitta,engels}
Note that in quantum wires with electron propagating along the x-axis $k_y$ and $k_z$ must be replaced by 
dynamical momentum operators.
The expectation value of  $k_y$, $k_z +e A_z/\hbar c$ with respect to the lowest subband state wave function of 
transverse degrees of freedom ($y,z$) vanish by symmetry considerations\cite{datta1990}
\be
\caH_{R} = \eta_R  k_x \sigma_z.
\ee

The \textit{bulk} Hamiltonian of Dresselhaus SOI is given by\cite{Rashba2005}
\be
\begin{split}
\caH_{\text{bulk},D} &= \gamma_c \Big[ \sigma_x k_x (k_y^2 - k_z^2) \\
&+\sigma_y k_y (k_z^2 - k_x^2) + \sigma_z k_z (k_x^2 -k_y^2) \Big ].
\end{split}
\ee
To obtain the effective Hamiltonian of quantum wire we have to take the average of the above bulk Hamiltonian
with respect to the ground state wave function of transverse ($y,z$) degrees of freedom.
In our geometry the Rashba electric field is applied in y-direction, and the lateral confining potential enforcing 
quasi one-dimensional motion is applied in z-direction.
Clearly $\la k_z \ra =0$ since the subband wavefunction along the z-axis has even parity. 
The subband wavefunction along the y-axis is a real function and therefore the expectation value  $\la k_y \ra=0$, too.
But we have to note that $\la y \ra \neq 0$ since the inversion symmetry is lacking in the y-direction.
The effective Hamiltonian for quantum wire is then
\be
\caH_{D} = \gamma_c \sigma_x k_x (\la k_y^2 \ra  - \la k_z^2 \ra)=\eta_D \sigma_x k_x,
\ee
where $\eta_D =\gamma_c (\la k_y^2 \ra  - \la k_z^2 \ra)$.

Now the one-particle Hamiltonian becomes
\be
\label{one-particle}
\caH_1 = E_y+E_z+ \frac{\hbar^2 k^2}{2 m^*} + \eta_R k \sigma_z + \eta_D k \sigma_x -E_Z \sigma_x.
\ee
The Dresselhaus term can be absorbed into the Zeeman term $E_Z = g_0 \mu_B B/2$ ($g_0 \approx 15$ for InAs)
in the following way.
\be
\e_Z \equiv E_Z - \eta_D k.
\ee
For the sake of completeness we include the Dresselhaus term in the calculation
of the band structure.  Later we will ignore it in the bosonization procedure.
By the diagonalization of the Hamiltonian Eq.~(\ref{one-particle}) the energy eigenvalues and
the corresponding normalized eigenvectors are obtained as follows:
For the lower band the eigenvalue is ($E_y,E_z$ put to zero)
\be
\begin{split}
E_-(k) &= \frac{\hbar^2 k^2}{2 m^*} - \sqrt{\e_Z^2 + \eta_R^2 k^2} \\
\end{split}
\ee
and the eigenvector is 
\be
\label{eigenvector1}
\xi_{-} =\begin{pmatrix} u_k^-  \cr v_k^- \cr \end{pmatrix},
\ee
where
\begin{align}
\label{eigenvector2}
u_k^-&=\frac{\e_Z}{\sqrt{(\eta_R k +D)^2 + \e_Z^2}}, \\
v_k^- &=\frac{\eta_R k + D }{\sqrt{(\eta_R k +D)^2 + \e_Z^2}},\\
\end{align}
Here
\be
D \equiv \sqrt{(\eta_R k)^2 + \e_Z^2}.
\ee
$u_k^-$ and $v_k^-$ represents the amplitudes for the spin to point in the  +z and -z direction, respectively.
For the upper band the results are given in  Ref. [\onlinecite{com}].


Quantum wires can be tailor made so that the quantities $\la k_y^2 \ra $ and $\la k_z^2 \ra$ are
{\it almost equal}.
If we assume the harmonic confining potential $ m^* \omega_0^2 z^2 /2$ along the z-axis 
we have $ \la k_z^2 \ra = m^* \omega_0 / 2 \hbar$.
For the y-direction  the constant  Rashba electric field  is acting so that the potential is linearly rising.
In this case \cite{sousa2003}
$\la k_y^2 \ra \sim 0.8 \left (\frac{ 2 m^* |e| E_0 }{\hbar^2} \right)^{2/3}.$
The condition $\la k_y^2 \ra=\la k_z^2 \ra $ is satisfied when the value of the electric field is given by
$eE_0 z_0=0.49\frac{\hbar^2}{2m^*}\frac{1}{z_0^2}$, where $z_0=\sqrt{\hbar/m^*\omega_0}$.
For this particular value of the electric field  the Dresselhaus term is negligible compared to the Zeeman energy 
and the Rashba coupling.
Note that
$E_-(k)$ becomes an even function of $k$ in this case.
Hereafter we assume this.
If the Rashba coupling becomes sufficiently strong such that
\be
 \eta_R^2 \ge \e_Z \hbar^2 /m^*
 \ee
then
the energy spectrum develops a double minium at 
\be
k=\pm \frac{1}{\eta_R} \Big [ \left( \frac{m^* \eta_R^2}{\hbar^2} \right )^2 -\e_Z^2 \Big]^{1/2}.
\ee
The energy at the minimum is given by
\be
E_{\text{min}} = - \frac{m^* \eta_R^2}{2 \hbar^2} - \frac{\hbar^2 \e_Z^2}{2 m^* \eta_R^2}.
\ee 
In such a case $E_{-}(0)>E_{\text{min}}$.
In our work we assume that $E_{-}(0)-E_{\text{min}}=-\epsilon_Z-E_{\text{min}}$ is less than the Fermi energy
so that there are only two Fermi wavevectors.

\section{Model for many-body Hamiltonian}
\label{interacting}
Let  $a_k$ and $b_k$ be the quasiparticle operators corresponding to  $E_-(k)$ and $E_+(k)$ , respectively.
They can be explicitly expressed in terms of electron operators as follows:
\begin{align}
\label{coefficients}
b_k^\dag&= c^\dag_{k \uparrow}  u_k^+ +  c^\dag_{k \downarrow}  v_k^+ ,\quad
a_k^\dag= c^\dag_{k \uparrow}  u_k^- +  c^\dag_{k \downarrow}  v_k^-, \nonumber \\
c^\dag_{k \uparrow}&= b_k^\dag u_k^+  +  a_k^\dag u_k^- ,\quad
c^\dag_{k \downarrow}= b_k^\dag v_k^+  +  a_k^\dag v_k^-.
\end{align}
When electrons are filled such that the Fermi energy is located in the energy gap between $a$ and $b$ bands
at $k=0$, we can safely neglect the $b$-type quasiparticles in the low energy regime.
Then the Eq.~(\ref{coefficients}) can be simplified.
\begin{align}
\label{simplified}
a_k^\dag &= c^\dag_{k \uparrow}  u_k^- +  c^\dag_{k \downarrow}  v_k^-, \nonumber \\
c^\dag_{k \uparrow}& \sim   a_k^\dag u_k^- ,\quad
c^\dag_{k \downarrow} \sim    a_k^\dag v_k^-.
\end{align}
A general electron-electron interaction in a quantum wire is given by
\be
\label{general}
\caH_{\text{int}} = \frac{1}{2} 
\sum_{k_1, k_2, q,  \sigma, \sigma'}
V_q c^\dag_{k_1 \sigma} c^\dag_{k_2 \sigma'} c_{k_2+q \sigma'} c_{k_1-q \sigma},
\ee
where $V_q$ is the interaction matrix element. Note that this interaction is spin-conserving.
For the long-range Coulomb interaction the interaction matrix element is
\be
V_q = \frac{2 e^2}{\e} K_0( |q| w)  \to \frac{2 e^2}{\e} \ln \frac{1}{|q| w} \quad \text{for} ~ |q| w \ll1.
\ee
$K_0$ is the modified Bessel function and 
$w$ is the cutoff length scale which is the order of the width of the quantum wire. 
$\e$ is the bulk dielectric constant.
For the short range interaction such as screened Coulomb interaction the matrix element $V_q$ can be taken to be
independent of the momentum transfer $q$.
Projecting the Hamiltonian Eq.~(\ref{general}) to the $a$-band with the use of Eq.~(\ref{simplified}),
we obtain
\be
\label{interaction}
\begin{split}
\caH_{\text{int}}  &=\frac{1}{2} \sum_{k_1,k_2,q}\,
\langle k_1, k_2 | \hat{V} | k_1-q, k_2+q \rangle\,  \\
& \times a^\dag_{k_1} a^\dag_{k_2}  a_{k_2+q}  a_{k_1-p},
\end{split}
\ee
where
\be
\label{projectedmatrix}
\begin{split}
&\langle k_1, k_2 | \hat{V} | k_1-q, k_2+q \rangle  \\
&= 
V_p \, [\xi_-^\dag(k_1) \xi_-(k_1-q)]
\, [\xi_-^\dag(k_2) \xi_-(k_2+q)]
\end{split}
\ee
is the projected interaction matrix element in the low energy Hilbert space.
The explicit expression of eigenvector $\xi_-$ is given by Eqs.~(\ref{eigenvector1},\ref{eigenvector2}).

At low energy, only the electron states near $-k_F$ and $k_F$ Fermi points 
need to be considered.
Following the usual procedures of g-ology and bosonization method\cite{voit}
we can express the interaction Hamiltonian Eq.~(\ref{interaction})
within g-ology scheme. 
Forward scattering $g_2$ and $g_4$ process.
Backscattering $g_1$ process
We note further  that for  fermions of a single species
 (like $a$-quasiparticle here)
$g_1$ process is identical with $g_2$ process.\cite{voit}
In this paper a commensurate filling is not considered, so that the Umklapp
processes ($g_3$) can be neglected. 
From now on we will call electrons with $k<0$ ($k>0$) left (right) movers. 


$\underline {g_4 ~\text{process}}$:
\begin{figure}[ht]
\begin{center}
\includegraphics[angle=0,width=0.8\linewidth]{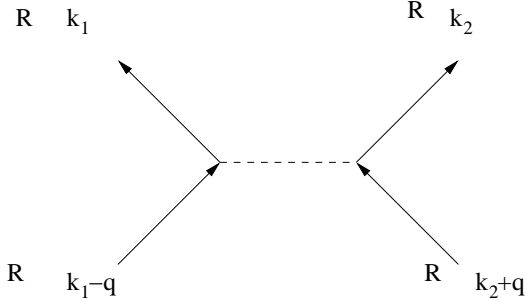}
\caption{A Feynman diagram of the $g_4$ process. All four momenta $k_1, k_2, k_2+q, k_1-q$ are located near the  right
Fermi point. The dotted line indicates the  matrix element $V_q$. See  text for details.}
\label{g4process}
\end{center}
\end{figure}
For instance let us assume that all four momenta $k_1, k_2, k_2+q, k_1-q$ are located near the right Fermi point in 
the  following. Then one can write $
k_i = +k_F + p_i$ with a condition $ \vert p_i \vert < \Lambda < k_F$.
$\Lambda$ is the momentum cutoff scale, within which the linearization of the $a$-band dispersion is valid.
It is convenient to introduce the right moving Dirac fermion operator $\psi_R$:  
$ \psi_R (p_i) \equiv a_{k_i}$  for $k_i = +k_F + p_i$.
We can make following approximation if we neglect  subleading contributions proportional to $(k-k_F)$ which are
irrelevant at low energy: $
[\xi_-^\dag(k_1) \xi_-(k_1-q)] \approx [\xi_-^\dag(k_F) \xi_-(k_F)] 
 = 1 $. 
Thus the effect of spin mixing reflected in the matrix elements $\xi_\sigma$ does not play any role for $g_4$ process.

The contributions from the neighborhood of left Fermi point can be treated in the same way.
The left moving  Dirac fermion operator $\psi_L$ can be introduced similarly.
$ \psi_L (p_i) \equiv a_{k_i}$ for $k_i = -k_F - p_i$.
The low energy effective Hamiltonian describing $g_4$ process can be read off from the original
Hamiltonian Eq.~(\ref{interaction}).
\be
\label{g4}
\caH_{g_4}=\frac{1}{2 N}\,\sum_{\vert q \vert < \Lambda} \, V_q \, \Big[ \rho_R(q) \rho_R(-q) + \rho_L(-q) \rho_L(q)\Big],
\ee
where $\rho_{R/L}(q)=\sum_p \psi^\dag_{R/L}(p+q) \psi_{R/L}(p)$ are the density operators of right and left moving Dirac
fermions. $N$ is the number of lattice sites of quantum wire. 
In the above expression the \textit{low-momentum asymptotics} of $V_q$ must be used.

$\underline{g_2 ~\text{ processes}}$:

\begin{figure}[ht]
\begin{center}
\includegraphics[angle=0,width=0.8\linewidth]{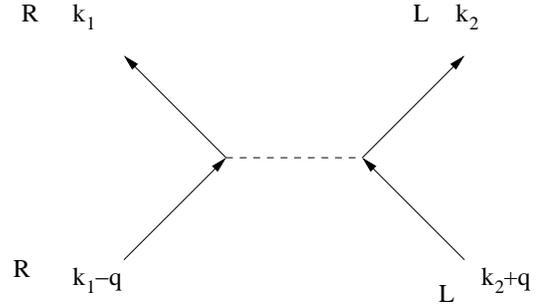}
\caption{A Feynman diagrams of the $g_2$ process. See text for details. There is another $g_2$ Feynman diagram in which 
$R \leftrightarrow L$.}
\label{g2process}
\end{center}
\end{figure}
According to the same reason as $g_4$ interaction
one can make the approximations in Fig. \ref{g2process}, 
\begin{equation*}
\begin{split}
[\xi_-^\dag(k_1) \xi_-(k_1-q)] &\approx [\xi_-^\dag(k_F) \xi_-(k_F)]=1 \\
[\xi_-^\dag(k_2) \xi_-(k_2+q)] &\approx [\xi_-^\dag(-k_F) \xi_-(-k_F)]=1 \\
\end{split}
\end{equation*}
Again the spin-mixing effect represented by the matrix elements does not modify the interaction.
Due to this  one can easily read off  the low energy effective Hamiltonian describing $g_2$ process from the original
Hamiltonian Eq.~(\ref{interaction}) using the definition of density operators $\rho_{R/L}(q)$.
\be
\label{g2}
\caH_{g_2} = \frac{1}{N}\,\sum_{\vert q \vert < \Lambda}\,
V_q\,\rho_R(q) \rho_L(-q) 
\ee

$\underline{g_{1,\parallel}~\text{ processes}}$:

\begin{figure}[ht]
\begin{center}
\includegraphics[angle=0,width=0.8\linewidth]{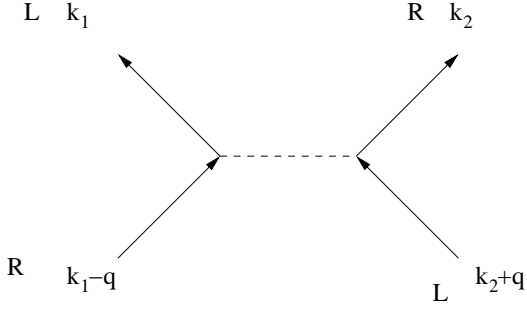}
\caption{A Feynman diagram of the $g_{1,\parallel}$ processes. There exists one more diagram where
$R \leftrightarrow L$. See text.}
\label{g1process}
\end{center}
\end{figure}
In this case the matrix elements play crucial role as can be seen in 
\be
\begin{split} 
&[\xi_-^\dag(k_1) \xi_-(k_1-q)] [\xi_-^\dag(k_2) \xi_-(k_2+q)] \\
& \approx[\xi_-^\dag(-k_F) \xi_-(+k_F)] [\xi_-^\dag(+k_F) \xi_-(-k_F)] \\
\end{split}
\ee
Evidently the dominant momentum transfer $q$ must be $2 k_F$. By changing the order of operators (thereby changing overall
sign of interaction) and by summing over momenta one arrives at
\be
\label{g1Hamiltonian}
H_{g_1} = -g V_{2 k_F} \int dx \rho_R(x) \rho_L(x),
\ee
where 
\be
\label{g-coupling}
\begin{split}
g &=\,[ \xi_-^\dag(-k_F)  \xi_-(+k_F) ] [\xi_-^\dag(+k_F)  \xi_-(-k_F) ] \\
&=  \frac{\e_Z^2}{\e_Z^2+(\eta_R k_F)^2}.
\end{split}
\ee
As can be seen from Eq.~(\ref{g-coupling}) the coupling constant $g$ depend on the applied \textit{magnetic field}
and the \textit{Rashba SOI} as well as Fermi momentum.

We observe that $g_2$ Hamiltonian Eq.~(\ref{g2}) and $g_{1}$ Hamiltonian Eq.~(\ref{g1Hamiltonian}) can be
combined completely. This is a special feature of fermions of \textit{single} species. In the presence of other
degrees of freedom such as spin a backscattering term ($g_{1,\perp}$ ) appears which is not of the Luttinger interaction
form.

\section{Bosonization and collective excitations}
The total effective Hamiltonian incorporating interaction is given by
\be
\label{total}
\caH = \caH^{(0)}+\caH_{g_4}+\caH_{g_2}+\caH_{g_1}.
\ee
The linearized non-interacting Hamiltonian $\caH_0$ is
\be
\caH^{(0)} =\sum_p \,\Big[ v_F p \psi^\dag_{R}(p) \psi_{R}(p)
- v_F p \psi^\dag_{L}(p) \psi_{L}(p) \Big ].
\ee
The Hamiltonian Eq.~(\ref{total}) can be bosonized straightforwardly.\cite{voit}
\begin{align}
\caH &= \pi v_F \int dx \,\Big[ : \rho_R^2(x) : +  : \rho_L^2(x) : \Big ] 
\nonumber \\
&+\frac{1}{2N}\,\sum_q\,V_q \,\Big[ \rho_R(q) \rho_R(-q) +  \rho_L(-q) \rho_L(q) \Big ]
\nonumber \\
&+\frac{1}{N} \sum_q \Big[ V_q - g V_{2 k_F} \Big ]\, \rho_R(q) \rho_L(-q).
\end{align}
$: :$ denotes normal ordering of operators.
It is convenient to introduce phase fields as follows:
\be
\label{phase}
\begin{split}
\theta(x) &= \frac{1}{2}\, \Big[ \phi_R(x) + \phi_L(x) \Big ],  \\
\phi(x) &= \frac{1}{2}\, \Big[ \phi_R(x) - \phi_L(x) \Big ], 
\end{split}
\ee
where $\rho_{R/L}(x) = \frac{1}{2\pi}\, \partial_x \phi_{R/L}(x)$.
In terms of phase fields
\begin{align}
\caH&= \frac{v_F}{2\pi}\,\int dx \,\Big[  (\partial_x \theta)^2 + (\partial_x \phi)^2 \Big ]
 \nonumber \\
&+\frac{1}{N}\, \sum_p\,\frac{ V_q q^2}{(2\pi)^2} \,
\Big[ \theta(q) \theta(-q) + \phi(q) \phi(-q) \Big ] \nonumber \\
&+ \left( \frac{1}{2\pi} \right )^2\frac{1}{N}\,\sum_p\,(V_q-g V_{2 k_F}) q^2 
\Big [ \theta(q) \theta(-q) -\phi(q) \phi(-q)  \Big]\nonumber \\
&=\frac{v_F}{2\pi}\,\int dx \,\Big[  (\partial_x \theta)^2 + (\partial_x \phi)^2 \Big ]
\nonumber \\
&+
\frac{1}{(2\pi)^2N}\,\sum_q \, 
\Big[ 2 (V_q-g V_{2 k_F}/2) q^2 \theta(q) \theta(-q)  \nonumber \\
&+ g V_{2 k_F} q^2 \phi(q) \phi(-q) \Big].
\end{align}
The Euclidean action is given by
\be
S[\theta,\phi]= \int d \tau 
\Big[ \int dx \, \frac{i}{\pi} \partial_x \phi \,\partial_\tau  \theta + H \Big].
\ee
In matrix form the above can be written as
\be
\begin{split}
\label{action}
S &=\frac{1}{2\pi}\,\int \frac{d \omega d q}{(2\pi)^2}\, 
\begin{bmatrix} \theta(-q,-\omega)  &  \phi(-q,-\omega) \cr \end{bmatrix} \nonumber \\
&\times
\begin{bmatrix} 
 v_\theta (q)  q^2   &  i q \omega  \cr
  i q \omega    &  v_\phi (q)  q^2  \cr 
  \end{bmatrix}
 \begin{bmatrix}   \theta(q,\omega)  \cr  \phi(q,\omega) \cr \end{bmatrix},
\end{split}
\ee
where
\begin{align}
\label{phase-velocities}
v_\theta &= v_\theta (q ) =  v_F \left(1 + \frac{V_q}{\pi v_F}-g\frac{ V_{2 k_F}}{2\pi v_F} \right),\nonumber \\
v_\phi &= v_\phi(q)=v_F \left(1 + \frac{g V_{2 k_F}}{2 \pi v_F} \right ).
\end{align}
$\theta$ and $\phi$ are the phase fields which are basically linear combination of density operators $\rho_{R/L}$ and 
they are  defined in Eq.~(\ref{phase}). 
 
The dispersion relation of the collective excitation can be obtained from the
kernel of action Eq.~(\ref{action}).
\be
\det \begin{bmatrix} 
 v_\theta (q)  q^2   &  i q \omega  \cr
  i q \omega    &  v_\phi (q)  q^2  \cr 
  \end{bmatrix} =0 
\ee
After analytic continuation $i\omega \to \omega$ we find 
\be
\label{dispersion1}
\omega = \big [v_\theta(q) v_\phi(q)\big ]^{1/2} q \equiv v_o q.
\ee
$v_0$ is the velocity of collective excitation. 
From Eq.~(\ref{phase-velocities}) one can write
\be
\label{velocity}
v_o = v_F \Big[ 1 + \frac{V_q}{\pi v_F} + \frac{(V_q-g V_{2 k_F}/2)(g V_{2 k_F}/2)}{(\pi v_F)^2} \Big ]^{1/2}.
\ee
The quantity in the bracket of Eq.~(\ref{velocity}) represents the renormalization effect due to
electron-electron interaction. The velocity of collective excitation can be controlled by
band filling, Rashba SOI, and magnetic field through dependence on $v_F$ and $g$.
Let us estimate the magnitude of the correction terms.   
In $v_\theta (q )$ the backscattering term, $g\frac{ V_{2 k_F}}{2\pi v_F}$, is a factor $g/2$ smaller than the 
forward term, $\frac{V_q}{\pi v_F}$.
In $v_\phi (q )$ the correction term  $gV_{2 k_F} / 2\pi v_F \sim 0.1g$ for the width of the quantum wire $w \sim100 \text{\AA}$
and $2 k_F \approx 1 \times 10^{6}  \text{cm}^{-1}$.
We also note that for the screened short range Coulomb interaction the interaction matrix element $V_q$ is almost
independent of momentum transfer $q$, and the backscattering term  plays an equally important role as forward
the scattering.

\section{Discussions and Summary}

It is instructive to compare this result with the velocities of phase fields  of collective excitation 
of ordinary Luttinger liquids.
For \textit{spinless fermions} it is given by
\begin{align}
v_\theta (q ) &=  v_F \left(1 + \frac{V_q}{\pi v_F}-\frac{ V_{2 k_F}}{2\pi v_F} \right),\nonumber \\
v_\phi (q ) &= v_F \left(1 + \frac{ V_{2 k_F}}{2 \pi v_F} \right ).
\end{align}

In Eq.~(\ref{velocities}) this corresponds to $g=1$, which implies absence of spin-orbit coupling 
and one type of spin, either up or down.
For the Luttinger liquids of \textit{spinful fermions} 
 the velocity of  charge  mode is  given  by
\begin{align}
v_{\theta_\rho} (q ) &=  v_F \left(1 + \frac{2V_q}{\pi v_F}-\frac{V_{2 k_F}}{2\pi v_F} \right),\nonumber \\
v_{\phi_\rho} (q ) &= v_F \left(1 + \frac{ V_{2 k_F}}{2 \pi v_F} \right ).
\end{align}

$\theta_\rho$ and $\phi_\rho$ are the phase fields in the charge sector.
 The spinful velocity
is recovered  with the replacement $V_q \rightarrow 2 V_q $ and $g=1$ in Eq.~(\ref{velocities}).
The velocity of the spin mode is 
\begin{align}
v_{\theta_s} (q ) &=  v_F \left(1 - \frac{V_{2 k_F}}{2\pi v_F} \right),\nonumber \\
v_{\phi_s}  (q )&= v_F \left(1 + \frac{ V_{2 k_F}}{2 \pi v_F} \right ).
\end{align}
$\theta_s$ and $\phi_s$ are the phase fields in the spin sector.
This corresponds to $V_q=0$ and $g=1$ in Eq.~(\ref{velocities}).

The dispersion relation of the collective mode may be measured by adding another quantum wire
parallel to the original wire in the presence of  a second magnetic field $\vec{B_t}$ along the y-axis.
When the first wire is located at $z=0$ and the second wire at $z=z_0$
the single particle energy dispersion of the second wire is $E(k)=\frac{\hbar^2(k-k_0)^2}{2m}$, where $k_0=eB_tz_0/\hbar c$,
m is the electron mass in the second wire,
and  the Landau gauge $\mathbf{A}_t=(z B_t,0,0)$ is used.
Wave-number selectivity due to momentum-resolved tunneling between
them, $E_{-} (k)=E(k)$, allows a mapping of the dispersion\cite{governale2002,boese2001,auslaender2002}.
Even in the presence of electron interactions this technique allows direct measurement of the collective excitation 
spectrum.\cite{Carpentier,Zulicke}

\begin{acknowledgments}
This work was  supported by grant No. R01-2005-000-10352-0 from the Basic Research Program 
of the Korea Science and Engineering
Foundation and by Quantum Functional Semiconductor Research Center (QSRC) at Dongguk University
of the Korea Science and Engineering
Foundation.  This work was completed during a  stay at KIAS.
\end{acknowledgments}



\begin{thebibliography}{}
\bibitem{com}For the upper band we have
$E_+(k) = \frac{\hbar^2 k^2}{2 m^*} +\sqrt{\e_Z^2 + \eta_R^2 k^2}$.
and $\xi_{+} =\begin{pmatrix} u_k^+  \cr v_k^+ \cr \end{pmatrix}$
where
$u_k^+=\frac{+(\eta_R k +D)}{\sqrt{(\eta_R k +D)^2 + \e_Z^2}}$ and $v_k^+ =\frac{-\e_Z}{\sqrt{(\eta_R k +D)^2 + \e_Z^2}}$.
\end{thebibliography}
\end{document}